\documentclass{basi}
\usepackage[british]{babel}
\usepackage[varg]{txfonts}
\usepackage{alltt}
\usepackage{rotating}
\usepackage{dcolumn}

\newcommand{\lapp}{\mbox{\raisebox{-0.3em}{$\stackrel{\textstyle <}{\sim}$}}}
\newcommand{\gapp}{\mbox{\raisebox{-0.3em}{$\stackrel{\textstyle >}{\sim}$}}}

\begin{document}
\title[A study of two asymmetric large radio galaxies]
      {A low-frequency study of two asymmetric large radio galaxies}
\author[A.~Pirya et~al.]%
       {A.~Pirya,$^1$\thanks{email: \texttt{akashpirya@gmail.com (AP);
            sumana@aries.res.in (SN), djs@ncra.tifr.res.in (DJS) and
            msingh@aries.res.in (MS)}}
       S.~Nandi,$^{1}$ D.~J.~Saikia$^2$ and M.~Singh$^{1}$\\
       $^1$Aryabhatta Research Institute of Observational Sciences (ARIES), Manora Peak, Nainital, 263 129, India\\
       $^2$National Centre for Radio Astrophysics, TIFR, Pune University Campus, Post Bag 3, Pune 411 007, India}

\pubyear{2011}
\volume{39}
\pagerange{\pageref{firstpage}--\pageref{lastpage}}

\date{Received 2011 December 27; accepted 2012 January 04}

\maketitle
%------------------------------------------------------------------------------%
% abstract and keywords                                                        %
%------------------------------------------------------------------------------%
\label{firstpage}

\begin{abstract}
We present the results of multifrequency observations of two asymmetric,
Mpc-scale radio sources with the Giant Metrewave Radio Telescope (GMRT) and 
the Very Large Array (VLA). The radio luminosity of these two sources,
J1211+743 and J1918+742, are in the Fanaroff-Riley class II (FRII) range, 
but have diffuse radio components on one side of the galaxy while
the opposite component appears edge-brightened with a prominent hot-spot. 
Although the absence of a hot-spot is reminiscent of FRI radio galaxies, 
suggesting a hybrid morphology, the radio jet facing the diffuse lobe
in J1211+743 is similar to those in FRII radio sources, and it
is important to consider these aspects as well while classifying these sources 
in the FR scheme. The observed asymmetries in these Mpc-scale sources are likely 
to be largely intrinsic rather than being due to the effects of orientation 
and relativistic motion. The formation of a diffuse lobe facing the radio 
jet in J1211+743 is possibly due to the jet being highly dissipative.
The low-frequency spectral indices of the lobes
are in the range of approximately $-$0.8 to $-$1, while at the outer edges these
vary from approximately $-$0.65 to $-$1.05 suggesting steep injection spectral indices, 
which need to be examined further from observations at even lower frequencies 
by telescopes such as the LOw Frequency ARray (LOFAR).  
\end{abstract}


%------------------------------------------------------------------------------%
% main text of the paper, using \section, \subsection, \subsubsection          %
%------------------------------------------------------------------------------%
\section{Introduction}\label{sec:Intro}
Studies of asymmetries in luminous extragalactic radio sources
have provided valuable insights towards understanding a number of
important aspects of these sources. While the correlation of depolarization
asymmetry with jet sidedness (Laing 1988; Garrington et al. 1988) has
established the importance of orientation in understanding jet asymmetry,
structural asymmetries of the oppositely-directed components have also been
used by a number of
authors over the years (e.g. Gilbert et al. 2004; Gopal-Krishna \& Wiita 2004;
Mullin, Riley \& Hardcastle 2008, and
references therein) to test the orientation-dependent unified schemes
for active galaxies since the early work by Kapahi \& Saikia (1982).
In addition to the effects of orientation, environmental asymmetries have
also been seen to play an important role. McCarthy, van Breugel \&
Kapahi (1991) showed the presence of more extended emission line gas
on the side of the source with the shorter component in radio galaxies.
In a sample of radio sources with or without radio jets, the shorter
component appears to depolarize faster (e.g. Pedelty et al. 1989) possibly due to
interaction with the gas clouds, two of the most extreme examples
being 3C254 (Thomasson, Saikia \& Muxlow 2006) and 3C459 (Thomasson,
Saikia \& Muxlow 2003). Radio structural asymmetries have been used to probe
intrinsic and environmental asymmetries on different scales ranging from the
compact steep-spectrum (CSS) and giga-Hertz peaked spectrum (GPS)
objects (e.g. Saikia et al. 1995, 2001) to the giant radio sources (GRS)
with projected sizes of over a Mpc (e.g. Subrahmanyan et al. 2008).
The possibility of the environment playing an important role in the
observed asymmetries is consistent with a slow velocity of propagation
of the jet head suggested by Scheuer (1995; see also O'Dea et al. 2009;
Konar et al. 2009). Structural asymmetries of the oppositely-directed 
components have been used to probe the `flip-flop'
model for radio sources, where energy supply from the nucleus is one-sided
at a time, but the source appears reasonably symmetric averaged over its
life time (e.g. Saikia \& Wiita 1982; Rudnick \& Edgar 1984).

Effects of orientation and relativistic motion on the
spectra of the hot-spots or lobes have also been studied (see Garrington,
Conway \& Leahy 1991; Dennett-Thorpe et al. 1997, 1999; Ishwara-Chandra
\& Saikia 2000; Mullin et al. 2008). While the high-brightness regions
appear to have flatter spectra on the side facing the jet, consistent with mild
relativistic beaming, the spectral asymmetries of the extended regions
appear to depend on the lobe length (Dennett-Thorpe et al. 1999). Liu
\& Pooley (1991a,b) found that for a sample consisting mostly of galaxies, the
more depolarized component has a steeper radio spectrum. Ishwara-Chandra
et al. (2001) examined the dependence of the Liu-Pooley relationship
on size,  optical identification and redshift and found it to be significantly
stronger for smaller sources, and similar for both radio galaxies and
quasars. In addition to Doppler effects, there also appears to be
intrinsic differences between the oppositely-directed components.

While most structural asymmetries can be understood by taking into account
the effects of both an asymmetric, inhomogeneous environment and relativistic
motion (e.g. Jeyakumar et al. 2005; see Gaibler, Khochfar \& Krause 2011
for results of recent simulations of jet-disk interaction),
MERLIN (Multi-Element Radio Linked Interferometer
Network) and VLA (Very Large Array) observations of one-sided radio sources
revealed a possible class of weak-cored, one-sided sources which were
difficult to understand  in the simple relativistic beaming scenario
(e.g. Saikia et al. 1989, 1990).  Saikia et al. (1996) also highlighted the
very asymmetric radio galaxy B0500+630, where one component had an FRII structure
(Fanaroff \& Riley 1974)  while the other one appeared to be of FRI type.
It was difficult to understand its asymmetries in the relativistic beaming
scenario.  Gopal-Krishna \& Wiita (2000) examined the detailed structure of
a large number of sources and identified several with FRI and FRII structures on
opposite sides, and christened these as HYMORS (HYbrid MOrphology Radio Sources).  
They inferred that such sources
indicate that the FR dichotomy is due to interaction with the external medium
rather than differences in the central engine.

%%%%%%%%%%%%%%%%%%%%%%%%%%%%%%%%%%%%%%%%%%%%%%%%%%%%%%%%%%%%%%%%%%%%%%%%%%%%%%%%%%%%
\begin{table}
\caption{Properties of the two sources.}
\label{tab:sample}
{\small
\centering
\begin{tabular}{c c c c c c c c c }
\hline
Source     &   RA (J2000) & Dec (J2000)   & $z$     & Refs. & LAS  &   size    & log P$^{1.4}_{t}$  \\
name       &  hh mm ss.ss & dd mm ss.s       &       &   & $^{\prime\prime}$  & kpc  & W/Hz   \\
(1)        &    (2)       &    (3)       &   (4)            &  (5)  & (6)  & (7)     & (8)     \\
\hline
J1211+743  &  12 11 59.03   & 74 19 04.7      & 0.1070 & MO79      & 437  &  846  & 25.17       \\
J1918+742  &  19 18 34.81   & 74 15 04.9      & 0.1940 & L01b;P    & 397  & 1266  & 25.63       \\
\hline

\end{tabular}

\footnotesize
References: MO79: Miley \& Osterbrock 1979; L01b: Lara et al. 2001b;
P: Present work (the position of the optical galaxy
was estimated from the Digital Sky Survey image). For J1211+743 and J1918+742, one arcsec 
corresponds to 1.935 and 3.190 kpc respectively. 
                        }
\end{table}
%%%%%%%%%%%%%%%%%%%%%%%%%%%%%%%%%%%%%%%%%%%%%%%%%%%%%%%%%%%%%%%%%%%%%%%%%%%%%%%%%%%%%%%%%%%%%

From a representative sample of $\sim$40 radio galaxies selected from the B2
sample, Parma et al. (1999) classified the spectral index variation along the oppositely-directed
components into three types. Those where the spectral index increases from the core outwards, which
is typical of the buoyant plumes of large FRI sources like 3C31 (Laing et al. 2008), were
called type 1, while those where the spectral index increases from the outer edges of
the components towards the core were called type 2. Type 2 sources could be of either FRI
or FRII type. In addition, a small fraction of $\sim$10 per cent showed no significant
variation along the source axis. These were christened as type 3, and non-variation of
spectral index along the components was suggested to be due to particle re-acceleration.
In a recent detailed study of lobe-like FRI radio galaxies which all had jets similar
to the ones in FRI sources embedded within their lobes, Laing et al. (2011) noted
similarities in the spectral index variation with FRII sources. While the jets and `caps'
at the edges of the lobes tend to have flatter spectra with spectral indices in the
range of $-$0.5--0.7, the outer edges of the lobes and plasma closer to the core tend to
have significantly steeper spectra. However, sometimes there is very little variation
over most of the lobe emission as seen in the images of B0755+37 and M84 by Laing et al. (2011),
the former being classified as type 3 by Parma et al. (1999).

\section{J1211+743 and J1918+742}
To further explore the asymmetries and low-frequency spectra and structure of 
Mpc-scale radio galaxies, we present the
results of Giant Metrewave Radio Telescope (GMRT) and VLA observations of two large, 
asymmetric radio sources,
J1211+743 and J1918+742, which have an FRII-like component containing a hot-spot
on one side and a diffuse lobe on the opposite side. In the case of J1211+743
a well-collimated radio jet faces the diffuse lobe.  These two sources are both 
associated with galaxies and have been selected
from the sample of large sources compiled by Lara et al. (2001a). Some of
their basic features are summarised in Table 1, which is arranged as follows. 
Column 1: Source name; 
columns 2 and 3: Right Ascension and Declination in J2000 co-ordinates;
column 4: redshift of the source; column 5: references for the position and redshift 
of the host galaxy; column 6: the largest angular size in arcsec;
column 7: linear size in kpc for a Universe with H$_o$=71 km s$^{-1}$ Mpc$^{-1}$, $\Omega_m$=0.27,
$\Omega_{vac}$=0.73 (Spergel et al. 2003);
column 8: the log of the luminosity (in W/Hz) at 1.4 GHz.

\subsection {Notes on the sources}
{\bf J1211+743 (4CT 74.17.01):} The radio galaxy J1211+743 with a projected
linear size of $\sim$846 kpc, has been observed by a number of authors over the
years (e.g. Rudnick \& Owen 1977; van Breugel \& Willis 1981; Fanti et al. 1983; J\"agers 1986; 
Zhao et al. 1989; Schoenmakers et al. 2000, 2001; Lara et al. 2001b). Rudnick
\& Owen (1977) identified the source with the south-west member of a pair of bright
galaxies in Abell 1500. The nearby companion galaxy is $\sim$7.5 arcsec away
(Lara et al. 2001b). The radio structure shows a prominent hot-spot in the southern
component while the northern component appears as a diffuse lobe (Figs. 1 and 2). The
maximum value of the peak brightness ratio in the oppositely-directed
components is $\sim$6.5. The source has a wiggling jet towards the north with the
magnetic field lines along the direction of the jet (van Breugel \& Willis 1981).
A weak counter-jet has been reported by Lara et al. (2001a).

\noindent
{\bf J1918+742:} The giant radio galaxy J1918+742 with a projected linear size of
$\sim$1.3 Mpc (Figs. 1 and 3) has been imaged earlier at 1.4 and 4.9 GHz with the VLA by
Lara et al. (2001a,b). It has a prominent hot-spot on the western component while the
eastern component is diffuse without a significant hot-spot. The maximum ratio of
the peaks of emission in the oppositely-directed components is $\sim$13.

The observations and analyses of these two sources are described briefly in Section 3,
while the observational results are presented in Section 4.
These results are discussed in Section 5, and summarised in Section 6.

%%%%%%%%%%%%%%%%%%%%%%%%%%%%%%%%%%%%%%%%%%%%%%%%%%%%%%%%%%%%%%%%%%%%%%%%%%%%%%%%%%%%
\begin{table}
\caption{Observing log.}
\label{tab:obs_log}
\begin{center}
{\small
\begin{tabular}{l l c c c c}
\hline
Source    & Telescope   & Obs. freq. &  Phase calibrator   & Obs. date  \\
          &             & MHz        &               &              \\
  (1)     &  (2)        & (3)        &  (4)          & (5)          \\
\hline
%---------------------------------------------------
J1211+743      &  GMRT      & 239  & J0834+555 & 2008 Dec 02 \\
               &            &      & J1459+716 &             \\
               &  GMRT      & 333  & J0834+555 & 2008 Dec 30 \\
               &            &      & J1459+716 &             \\
               &  GMRT      & 607  & J0834+555 & 2008 Dec 02 \\
               &            &      & J1459+716 &             \\
               &  VLA$^a$   & 4885 & J1048+717 & 1997 Jul 27 \\
\\
%--------------------------------------------
J1918+742      & GMRT       & 239  & J1459+716 & 2008 Dec 27 \\
               &            &      & J2350+646 &             \\
               & GMRT       & 332  & J1459+716 & 2009 Mar 10 \\
               & GMRT       & 607  & J1927+739 & 2009 Dec 27 \\
\hline

\end{tabular}

                }
\footnotesize
$^a$ archival data from the VLA \\
\end{center}
\end{table}
%%%%%%%%%%%%%%%%%%%%%%%%%%%%%%%%%%%%%%%%%%%%%%%%%%%%%%%%%%%%%%%%%%%

%%%%%%%%%%%%%%%%%%%%%%%%%%%%%%%%%%%%%%%%%%%%%%%%%%%%%%%%%%%%%%%%%%%%%%%%%%%%%%%%%%%%
\section{Observations and analyses}
\label{sec:obs}
The GMRT and VLA observations were made in the standard fashion,
with the target source observations being interspersed by the
observations of the phase calibrators.
The observing log for all the observations is shown in Table 2
which is arranged as follows.
Columns 1 and 2: name of the source and the telescope;
column 3: the frequency of the observations in MHz;
column 4: phase calibrators used for the different observations;
column 5: the dates of the observations.
The primary flux density and bandpass calibrators
were 3C48, 3C147 and 3C286. 
The flux densities at frequencies below 408 MHz have been
extrapolated using the parameters given by Baars et al. (1977),
while at higher frequencies these are in the Baars et al. (1977) scale.
At low frequencies, the assumed flux densities of 3C48 are 50.72 
and 43.41 Jy at 239 and 332 MHz respectively, while for 3C286 the
corresponding values are 28.07 and 25.96 Jy respectively. 3C147 was
observed at 333 MHz and its flux density was assumed to be 52.69 Jy.
The total observing time on each source is about 8 hr for the
GMRT observations and $\sim$10 min for the VLA observations.
All the data were analyzed in the standard
fashion using the NRAO {\tt AIPS} package. The low-frequency GMRT data were
significantly affected by radio frequency interference (RFI),
and these data were flagged using different tasks, such as {\tt UVFLG, TVFLG}
and {\tt SPFLG} available in {\tt AIPS}. For
the GMRT observations, we initially calibrated the data for one channel, and
then applied bandpass calibration, using the task {\tt BPASS} for calculating
the gains of all other channels. Channel averaging was done to obtain the continuum
data sets. These continuum data sets were imaged and CLEANed using multiple facets for
the different low-frequency GMRT observations using the task {\tt IMAGR}. All the
data were self calibrated using a clean component model in the task {\tt CALIB}
to produce the final images, which were then corrected for the primary beam response.

%%%%%%%%%%%%%%%%%%%%%%%%%%%%%%%%%%%%%%%%%%%%%%%%%%%%%%%%%%%%%%%%%%%%%%%%%%%%%%%%%%%%%
\begin{figure}
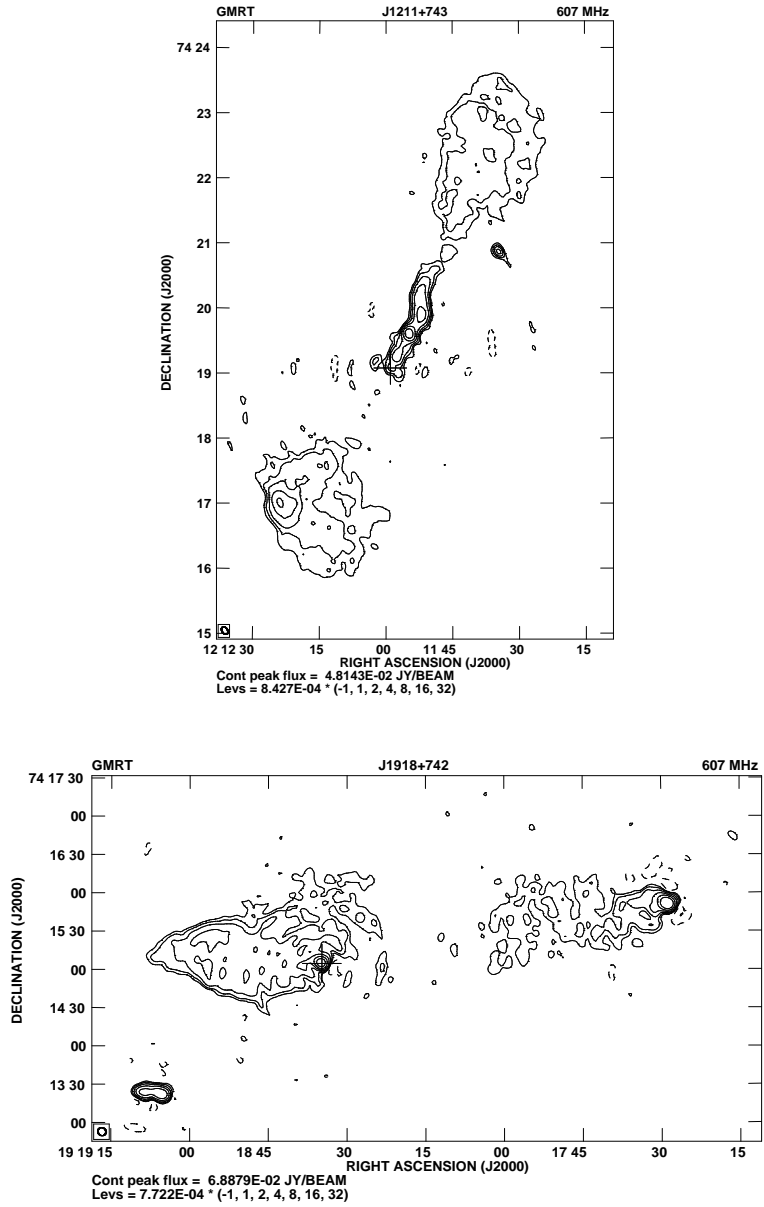

\centerline{\includegraphics[width=6.3cm]{J1211_610_N3.PS}} %\qquad
\centerline{\includegraphics[width=6.3cm,angle=-90]{J1918_610_N3.PS}}
\caption{GMRT full-resolution images of J1211+743 and J1918+742 at 607 MHz with
angular resolutions of $\sim$6 arcsec, which illustrate the diffuse lobes
without a prominent hot-spot, reminiscent of FRI lobes, on one side, and an
edge-brightened lobe with a hot-spot on the opposite side. The $+$ sign 
denotes the position of the optical galaxy.}
\end{figure}
%%%%%%%%%%%%%%%%%%%%%%%%%%%%%%%%%%%%%%%%%%%%%%%%%%%%%%%%%%%%%%%%%%%%%%%%%%%%%%%%%%%%%

%%%%%%%%%%%%%%%%%%%%%%%%%%%%%%%%%%%%%%%%%%%%%%%%%%%%%%%%%%%%%%%%%%%%%%%%%%%%%%%%%%%%%

\section{Observational results}
\label{sec:Obs_res}
The GMRT full-resolution images of the J1211+743 and J1918+742 with angular resolutions of
7.6$\times$5.1 arcsec$^2$ ~along position angle (PA) 32$^\circ$ and 6.6$\times$6.1 arcsec$^2$
along ~PA 42$^\circ$ are presented in Fig. 1. The peak brightness of the northern lobe of
J1211+743 is 3 mJy/beam compared with 15 mJy/beam for the southern lobe.
The corresponding values for the western and eastern
lobes of J1918+742 are 69 and 7 mJy/beam respectively.  The angular sizes of the lobes are smaller
than the largest scale structures visible to the GMRT (http://gmrt.ncra.tifr.res.in),
and we have checked that short-spacing data were available to image these sources reliably.
To study the spectra of the lobes over at least a factor of $\gapp$5 in frequency the images
of J1211+743 have been convolved with an angular resolution of 18 arcsec, similar
to the coarsest resolution of our images. The images at frequencies $>$300 MHz were initially
made by tapering the data and then convolving these to a common resolution.  For
J1918+742 we have chosen an angular resolution of 45 arcsec so that these are
all similar to the NRAO VLA Sky Survey (NVSS; Condon et al. 1998) image. 
However, we also imaged the GMRT low-frequency data of J1918+742 with an angular resolution of 
16 arcsec over a smaller frequency range to check for consistency of the results.
The 18-arcsec resolution images of J1211+743 and the 45-arcsec resolution images of J1918+742
are shown in Figs. 2 and 3 respectively, while
the observational parameters and some of the observed properties are tabulated in Table 3
which is arranged as follows.
Column 1: Frequency of the observations in units of MHz, where G or V indicates either GMRT or
VLA observations respectively; column 2: the resolution of the image in arcsec;
column 3: the rms noise in units of mJy/beam; columns 4, 5: peak and integrated flux density
of the entire source in units of mJy/beam and mJy respectively.
The flux densities for each source and its components at different frequencies have been estimated
over similar areas. Columns 6, 9, 11: components designations, where N, W, C, S, E denote northern,
western, core, southern and eastern components respectively; columns 7 and 8, 12 and 13:
peak and total flux densities of each component in units of mJy/beam and mJy respectively. For the cores, only
the peak brightness is quoted in column 10, to minimise contamination by extended emission in their vicinity.
The total flux densities have been estimated by specifying an area around the components.
The components refer to the entire lobe emission, but excluding the radio jet in J1211+743 whose
flux densities are listed at the bottom of Table 3.
We examined the change in flux density by varying the area around the components, and
find that the typical errors in the flux densities, including calibration errors, are
$\sim$14  per cent at $\sim$235 and 325 MHz, $\sim$10 per cent at 610 MHz and $\sim$5 per cent at
higher frequencies (cf. Konar et al. 2008; Nandi et al. 2010). The integrated spectra
of the sources, as well as the spectra of the different components by combining the results
of our observations with existing data in the literature are shown in the left panels of
Figs. 4 and 5.
We have fitted the spectra using both a linear least squares fit as well as a parabolic fit, and
find the linear fits to be better. 

%%%%%%%%%%%%%%%%%%%%%%%%%%%%%%%%%%%%%%%%%%%%%%%%%%%%%%%%%%%%%%%%%%%%%%%%%%%%%%%%%%%%%%%%%%%%%%%%%%
%%%%%%%%%%%%%%%%%%%%%%%%%%%%%%%%%%%%%%%%%%%%%%%%%%%%%%%%%%%%%%%%%%%%%%%%%%%%%%%%%%%%%%%%%%%%%%%%%%
\begin{table}
\caption{The observational parameters and observed properties of the sources.}
\label{tab:obs_prop}
{\small
\centering
\begin{tabular}{l r c r r c r r c r c r r}
\hline
Freq. & {res.} &rms & S$_{p}$ & S$_{I}$ & Cmp. & S$_{p}$ & S$_{t}$ & Cmp. & S$_{p}$ & Cmp. & S$_{p}$ & S$_{t}$ \\
MHz  & $^{\prime\prime}$ & mJy & mJy & mJy &  & mJy & mJy &  & mJy  &  & mJy & mJy    \\
     &                   & /b  & /b  &     &  & /b  &     &  & /b      &  & /b  &        \\
(1)  &  (2)              & (3) & (4) & (5) & (6) & (7) & (8) & (9)  & (10) & (11) & (12) & (13) \\
\hline
                                       \multicolumn{13}{c}{{\bf  J1211+743$^{\star}$}} \\
\\
G239  &  18 & 1.48  &218  & 3681  & N &49  & 1359   & C & $\leq$32       & S    &163  &1533  \\
G333  &  18 & 0.81  &168  & 2507  & N &33  &  877   & C & $\leq$16       & S    &118  &1033  \\
G607  &  18 & 0.79  &104  & 1420  & N &23  &  530   & C & 11             & S    & 73  & 595  \\
V4885 &  18 & 0.27  & 26  &  234  & N & 3  &   49   & C & 13             & S    & 18  &  96  \\
\\
\multicolumn{13}{c}{{\bf  J1918+742}} \\
\\
G239  &  16 & 1.81  &194  & 2500  & W &194 & 675    & C & $\leq$70       & E    & 115  & 1760 \\
      &  45 & 5.98  &481  & 2575  & W &311 & 691    & C &                & E    & 481  & 1843 \\
G332  &  16 & 0.76  &175  & 2240  & W &175 & 627    & C & $\leq$64       & E    & 114  & 1570 \\
      &  45 & 3.07  &416  & 2308  & W &267 & 661    & C &                & E    & 416  & 1648 \\
G607  &  16 & 0.69  &102  & 1089  & W &102 & 319    & C & 46             & E    &  61  &  608  \\
      &  45 & 1.80  &207  & 1176  & W &140 & 348    & C &                & E    & 207  &  741  \\
V1400 &  45 & 0.89  &127  &  575  & W & 88 & 175    & C &                & E    & 127  &  400  \\
\\
\hline
\end{tabular}
\footnotesize
$^\ast$ The peak and total flux densities of the jet from the images with an angular resolution of
         18 arcsec are as follows. 239 MHz: 218  mJy/b and 819 mJy;
        333 MHz: 168 mJy/b and 549 mJy; 607 MHz: 104 mJy/b and 296 mJy; 4885 MHz: 26 mJy/b and 77 mJy
            }
\end{table}

%%%%%%%%%%%%%%%%%%%%%%%%%%%%%%%%%%%%%%%%%%%%%%%%%%%%%%%%%%%%%%%%%%%%%%%%%%%%%%%%%%%%%

%%%%%%%%%%%%%%%%%%%%%%%%%%%%%%%%%%%%%%%%%%%%%%%%%%%%%%%%%%%%%%%%%%%%%%%%%%%%%%%%%%%%%
\begin{figure}
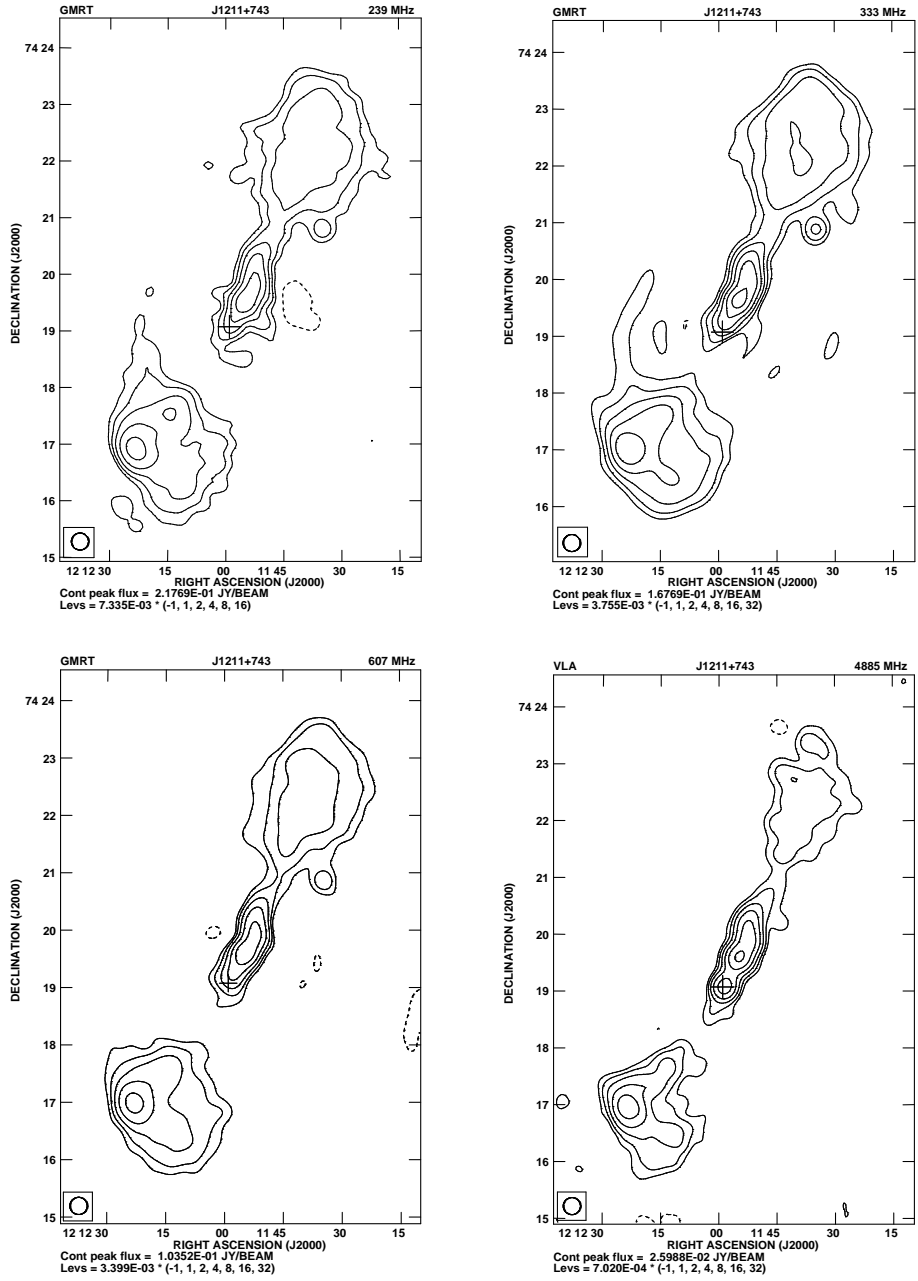

\centerline{\includegraphics[width=5.7cm]{1211_235_LOW_KNTR.PS} \qquad
            \includegraphics[width=5.7cm]{1211_325_LOW_KNTR.PS}}
\medskip
\centerline{\includegraphics[angle=0,width=5.7cm]{1211_610_LOW_KNTR.PS} \qquad
            \includegraphics[origin=lt,angle=0,width=5.7cm]{1211_C_LOW_KNTR.PS}}
\caption{GMRT low-frequency images of J1211+743 at 239, 333, 607 MHz and
the VLA high-frequency image at 4885 MHz with an angular resolution of 18 arcsec.
The $+$ sign indicates the position of the optical galaxy. }
\end{figure}
%%%%%%%%%%%%%%%%%%%%%%%%%%%%%%%%%%%%%%%%%%%%%%%%%%%%%%%%%%%%%%%%%%%%%%%%%%%%%%%%%%%%%
%%%%%%%%%%%%%%%%%%%%%%%%%%%%%%%%%%%%%%%%%%%%%%%%%%%%%%%%%%%%%%%%%%%%%%%%%%%%%%%%%%%%%
\begin{sidewaysfigure}
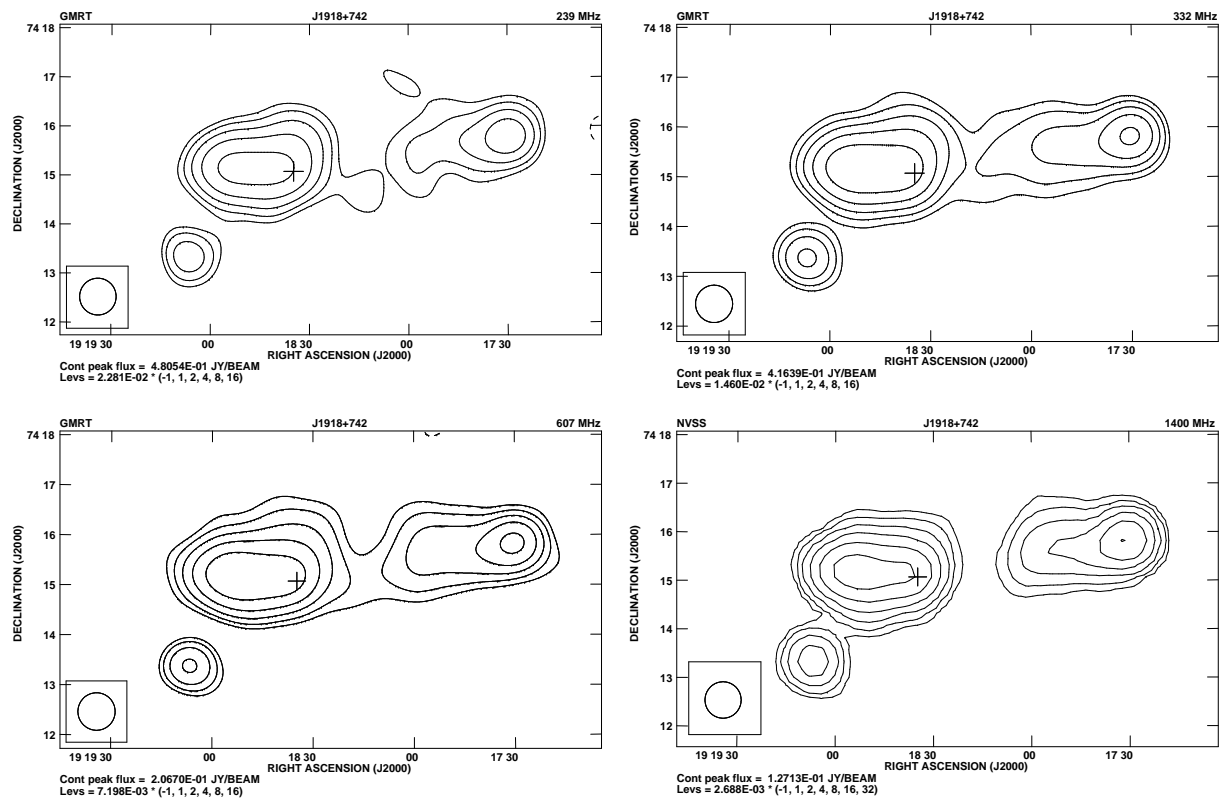

\centerline{\includegraphics[angle=-90,width=8.09cm]{1918_235_LOW_NVSS.PS} %\qquad
            \includegraphics[angle=-90,width=8.09cm]{1918_325_LOW_NVSS.PS}} 
%\centerline{\includegraphics[angle=-90,width=8.09cm]{1918_325_LOW_NVSS.PS}} 

%\medskip
\centerline{\includegraphics[angle=-90,width=8.09cm]{1918_610_LOW_NVSS.PS} % \qquad
            \includegraphics[angle=-90,width=8.09cm]{1918_NVSS_LOW_NVSS.PS}}
%           \includegraphics[origin=lt,angle=-90,width=8.09cm]{1918_NVSS_LOW_NVSS.PS}}
%\centerline{\includegraphics[origin=lt,angle=-90,width=8.09cm]{1918_NVSS_LOW_NVSS.PS}}
%
%\vspace{-3cm}
\caption{GMRT low-frequency images of J1918+742 at 239, 332 and 607 MHz, and the VLA
NVSS image at 1400 MHz. The resolution of these images is 45 arcsec.  
The $+$ sign indicates the position of the optical galaxy. }
\end{sidewaysfigure}
%%%%%%%%%%%%%%%%%%%%%%%%%%%%%%%%%%%%%%%%%%%%%%%%%%%%%%%%%%%%%%%%%%%%%%%%%%%%%%%%%%%%%

%%%%%%%%%%%%%%%%%%%%%%%%%%%%%%%%%%%%%%%%%%%%%%%%%%%%%%%%%%%%%%%%%%%%%%%%%%%%%%%%%%%%%%%%%%%%%%%%%%
\section{Results and discussions}
\label{sec:res_and_dis}
\subsection{Component brightness asymmetry}
\label{sec:res_and_dis:compt_bright_asymmtery}
Although the radio luminosity of these sources is in the FRII category, one of the
components in these two sources is diffuse with no
significant hot-spot at its outer edge. Although the diffuse lobes are reminiscent
of those in FRI radio galaxies, the radio jet in J1211+743 is well-collimated (see Fig. 1)
with the magnetic field lines following the bends in the jet (van Breugel \& Willis 1981).
This is similar to what is usually seen in jets in FRII radio sources (e.g. Bridle \&
Perley 1984), suggesting that one needs to examine the jet and magnetic field structures
as well as the structure of the lobes before classifying these sources as being of 
hybrid FRI/FRII morphology. 

Asymmetries in the peak brightness of the lobes
could be due to either relativistic motion of the hot-spots or intrinsic asymmetries
in the sources or their environments. However, since these are all
large Mpc-scale sources associated with galaxies, relativistic motion is not likely 
to be a dominant factor. The fraction of emission from the core at an emitted frequency
of 8 GHz, which is a statistical indicator of orientation of the jet axis to the line of
sight is $\sim$0.08 and 0.05 for J1211+743 and J1918+742 respectively. These values overlap
with the distribution for quasars and are significantly higher than the median value of 0.002
for 3CRR radio galaxies (Saikia \& Kulkarni 1994). We therefore assume that these are
close to the dividing line between galaxies and quasars and adopt an inclination angle of 
45$^\circ$ (Barthel 1989). Most luminous radio galaxies are likely to have small values
of speeds of advancement of $\lapp$0.1$c$  (e.g. Scheuer 1995; see also O'Dea et al. 2009; 
Konar et al. 2009). For J1211+743, Machalski et al. (2007) estimated an advancement speed
of the head of 0.009$c$ from their modelling of the dynamics of the source.
The maximum values of the peak brightness ratio on opposite sides of the nucleus are
$\sim$6.5 and 13 for the two sources respectively (see Table 3). For an inclination angle, $\phi$, of 
45$^\circ$, an advancement speed ($\beta$$c$) of 0.1$c$, and a spectral index,
$\alpha\sim$$-$1, (defined as S$\propto\nu^{\alpha}$),
the expected brightness ratio, {$\{(1+\beta~{\rm cos}\phi)/(1-\beta~{\rm cos}\phi)\}^{(2-\alpha)}$}
(e.g. Blandford \& K\"onigl 1979)  is only $\sim$1.5. This is significantly smaller than the
observed values suggesting intrinsic asymmetries in the sources. In the case of J1211+743, the diffuse
lobe is on the side of the radio jet, which is also  contrary to what would be expected in the relativistic
beaming scenario. The ratio of the peak brightness in the jet of J1211+743 to 3$\sigma$, where 
$\sigma$ is the rms noise in the counter-jet side in the GMRT 607-MHz image, is $\gapp$50. 
This implies that the jet velocity
would have to be $\gapp$0.8$c$ for $\phi$$\sim$45$^\circ$ and $\alpha\sim$$-$0.7, if the asymmetry
is due to bulk relativistic motion. Considering the overall structure of the source, the 
observed jet asymmetry could be largely due to an asymmetric dissipation of energy on the opposite 
sides, leading to the formation of a diffuse lobe without a prominent hot-spot on the jet side.  
Such a  possibility has been suggested earlier for a number of radio jets in quasars such as in
B1004+13, B1857+566 and 3C280.1 where the radio jet points towards a lobe without a significant 
hot-spot, while the edge-brightened component on the opposite side has a prominent hot-spot 
(e.g. Saikia 1984; see also Gopal-Krishna, Wiita \& Hooda 1996).

\subsection{Radio spectra}
\label{sec:res_and_dis:rad_spec}
The integrated radio spectra and those of the components for J1211+743 and J1918+742
are shown in Figs. 4 and 5 respectively. For J1211+743, the integrated spectral index
is $-$0.83$\pm$0.02, while the values of $\alpha$ for the northern, southern and jet
components are $-$1.02$\pm$0.03, $-$0.88$\pm$0.02 and $-$0.71$\pm$0.03 respectively.
The integrated spectral index for J1918+742 is $-$0.97$\pm$0.04, while the spectral
indices for the western and eastern components are $-$0.77$\pm$0.05 and $-$1.05$\pm$0.06
respectively. We attempted both a straight-line and parabolic fit, and find that straight
line fits are consistent with the available data, although for J1211+743 the 10.7 GHz measurements
by Schoenmakers et al. (2000, 2001) suggest a steepening of the spectra of the lobes. As expected, 
both the components with a hot-spot have flatter spectral indices than the components 
on the opposite side. 

We also examine the variation in spectral index across the components for the two sources.
To get reliable values of spectral index profiles we determined the spectral indices
from measurements at a minimum of four 
frequencies differing by a factor of at least $\sim$5. High-frequency 
measurements at greater than a few GHz were available only for J1211+743. As discussed
earlier, the images of J1211+743 were made with an angular resolution of 
18 arcsec, similar to the lowest resolution of our observations. For J1918+742 
the low-frequency GMRT images were convolved with a resolution of 45 arcsec, the resolution of the
NVSS image, since higher frequency data were not available. For each source we 
estimated the spectral indices in identical areas with significant flux density at all the frequencies, 
and separated by about a beamwidth. 
The images were aligned, rotated to be along the east-west direction and regridded
so that the flux densities in each area of a source were estimated in identical boxes at
the different frequencies using the {\tt AIPS} task
{\tt IMEAN}. We have followed a similar procedure as in our earlier studies
(e.g. Jamrozy et al. 2008; Konar et al. 2008; Nandi et al. 2010). 
The right panels of Figs. 4 and 5 show the spectral index variation across the components.
Distances have been measured from either the hot-spot or the centre of the first
strip at the outer edge of the component.   

For the radio galaxy J1211+743, there have been a couple of attempts in the past to 
examine the variation of spectral index across the source from observations at a limited
number of frequencies. J\"agers (1986) found the two-point spectral index $\alpha_{609}^{1415}$
to exhibit a large steepening with distance from the hot-spot for the southern component with 
$\alpha$ varying from approximately $-$0.7 to $-$2.0 over a distance of $\sim$2 arcmin ($\sim$230 kpc), 
a relatively smaller gradient for the northern diffuse lobe with $\alpha$ varying 
from approximately $-$0.8 to $-$1.2 over a similar length scale, and the jet region to show very
little variation along its length with $\alpha$ approximately $-$0.7. 
Schoenmakers et al. (2000, 2001) also find a flatter spectrum region in the vicinity of the radio
jet in both the spectral index profiles $\alpha_{354}^{10450}$  and $\alpha_{354}^{1400}$
with $\alpha$$\sim$$-$0.7 to $-$0.65. However their profiles
$\alpha_{354}^{10450}$ and $\alpha_{354}^{1400}$ for both the lobes are very different from 
each other. For
example, $\alpha_{354}^{10450}$ appears to flatten farther away from the edge and towards the core
for the northern lobe and exhibits no significant variation in the southern lobe, while 
$\alpha_{354}^{1400}$ appears to steepen with distance from the edge for both the lobes.  
Over the limited distances that we have determined the profiles for $\alpha_{239}^{4885}$,
which have been estimated by a least-squares fit to flux densities at four frequencies
in regions of high signal to noise ratio, the trends are broadly consistent with those of
J\"agers (1986). The southern lobe exhibits a significant steepening with $\alpha$ varying
from approximately $-$0.8 to $-$1.1 over a distance of $\sim$50 arcsec ($\sim$100 kpc) while the
variation in the northern lobe is very marginal over a similar distance. 
For the radio galaxy J1918+742, where the eastern component is diffuse with no prominent hot-spot,
and the optical galaxy/radio core is shifted significantly towards the east,
the spectra of both components appear to steepen with distance from the outer edge of the components.
Here the spectral indices $\alpha_{239}^{1400}$ have been determined from images with an angular
resolution of 45 arcsec, and the spectral index varies from approximately $-$0.65 to $-$1.05 over 
$\sim$120 arcsec ($\sim$380 kpc) for the western lobe and from approximately $-$0.75 to $-$0.95 over 
$\sim$80 arcsec ($\sim$255 kpc). The trend is consistent with what is normally observed in 
most FRII radio galaxies. 

%%%%%%%%%%%%%%%%%%%%%%%%%%%%%%%%%%%%%%%%%%%%%%%%%%%%%%%%%%%%%%%%%%%%%%%%%%%%%%%%%%%%%%%%%%%%%%%%%%%%%%%%%%
\begin{figure}
\centerline{\includegraphics[width=6.2cm]{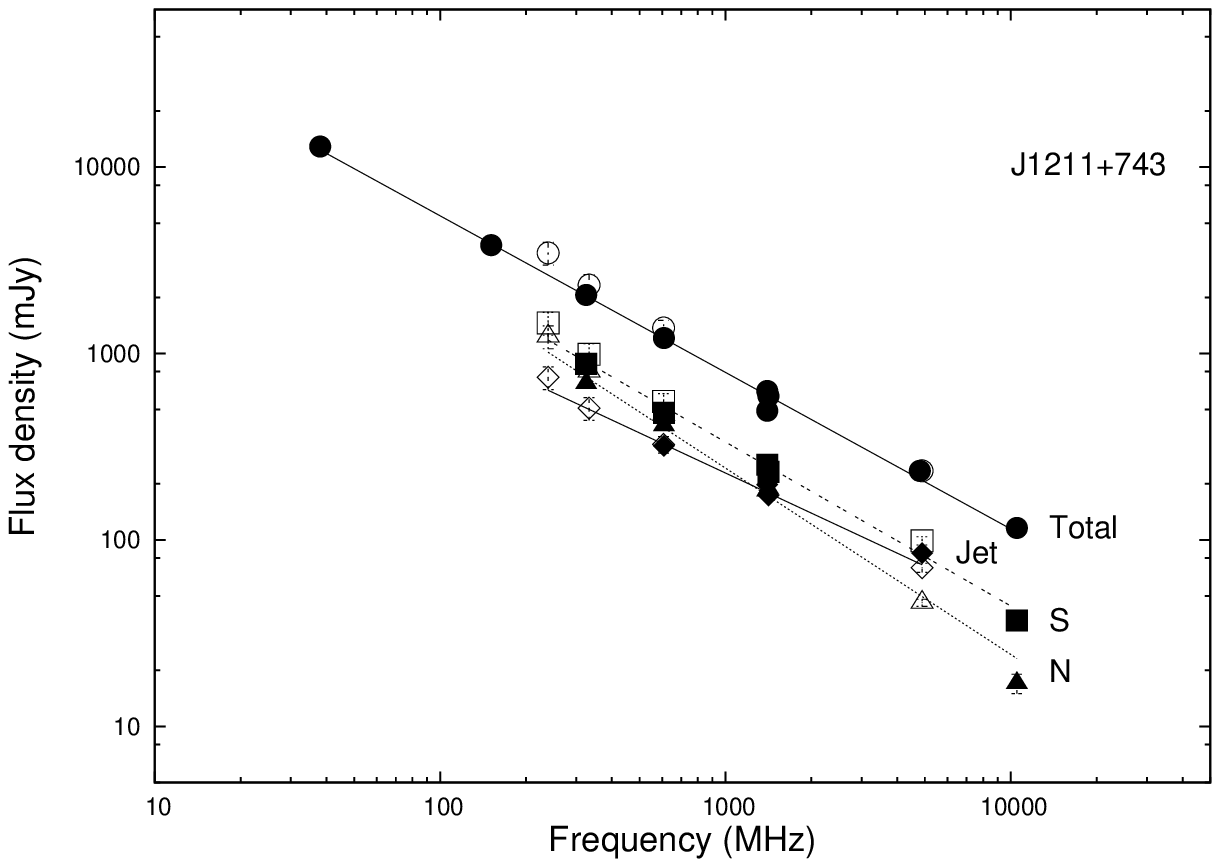} \qquad
            \includegraphics[width=6.2cm]{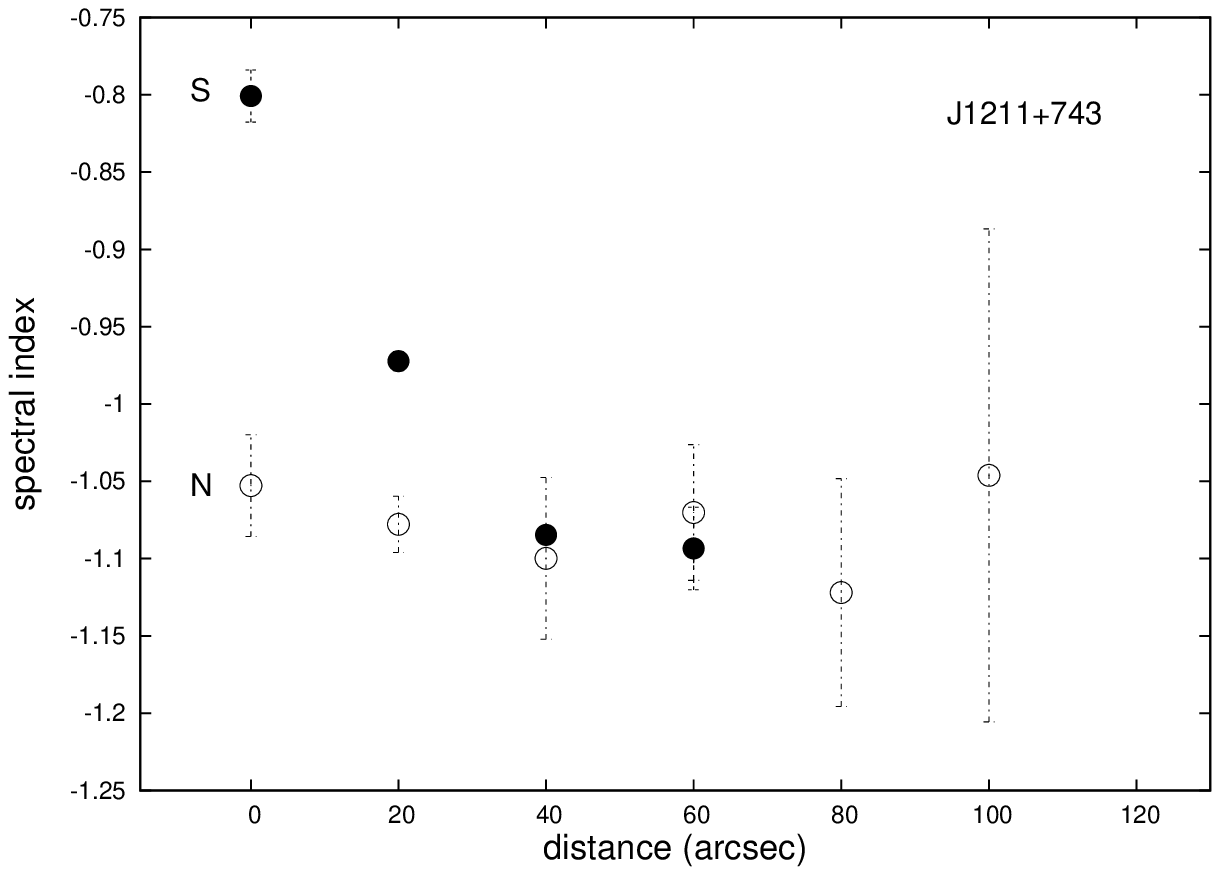}}
\caption{ Left panel: the spectra of the J1211+743  and its components.
          The values have been taken from the following references: 38 MHz, Rees (1990); 151 MHz, Hales et al. (1991);
          325, 4850 and 10500 MHz, Schoenmakers et al. (2000, 2001); 608.5, 1415 and 4885 MHz,
          van Breugel \& Willis (1981); 1400 MHz, NVSS survey; 1400 MHz, White \& Becker (1992); 239, 333, 607 and
          4885 MHz, from this paper.
          Right panel: The spectral index profiles for the northern and southern components with an angular
          resolution of 18 arcsec, using our measurements at 239, 333, 607 and 4865 MHz. For J1211+743, one arcsec
          corresponds to 1.935 kpc.
        }
\end{figure}
%%%%%%%%%%%%%%%%%%%%%%%%%%%%%%%%%%%%%%%%%%%%%%%%%%%%%%%%%%%%%%%%%%%%%%%%%%%%%%%%%%%%%%%%%%%%%%%%%%%%%%%%%%
%%%%%%%%%%%%%%%%%%%%%%%%%%%%%%%%%%%%%%%%%%%%%%%%%%%%%%%%%%%%%%%%%%%%%%%%%%%%%%%%%%%%%%%%%%%%%%%%%%%%%%%%%%
\begin{figure}
\centerline{\includegraphics[width=6.2cm]{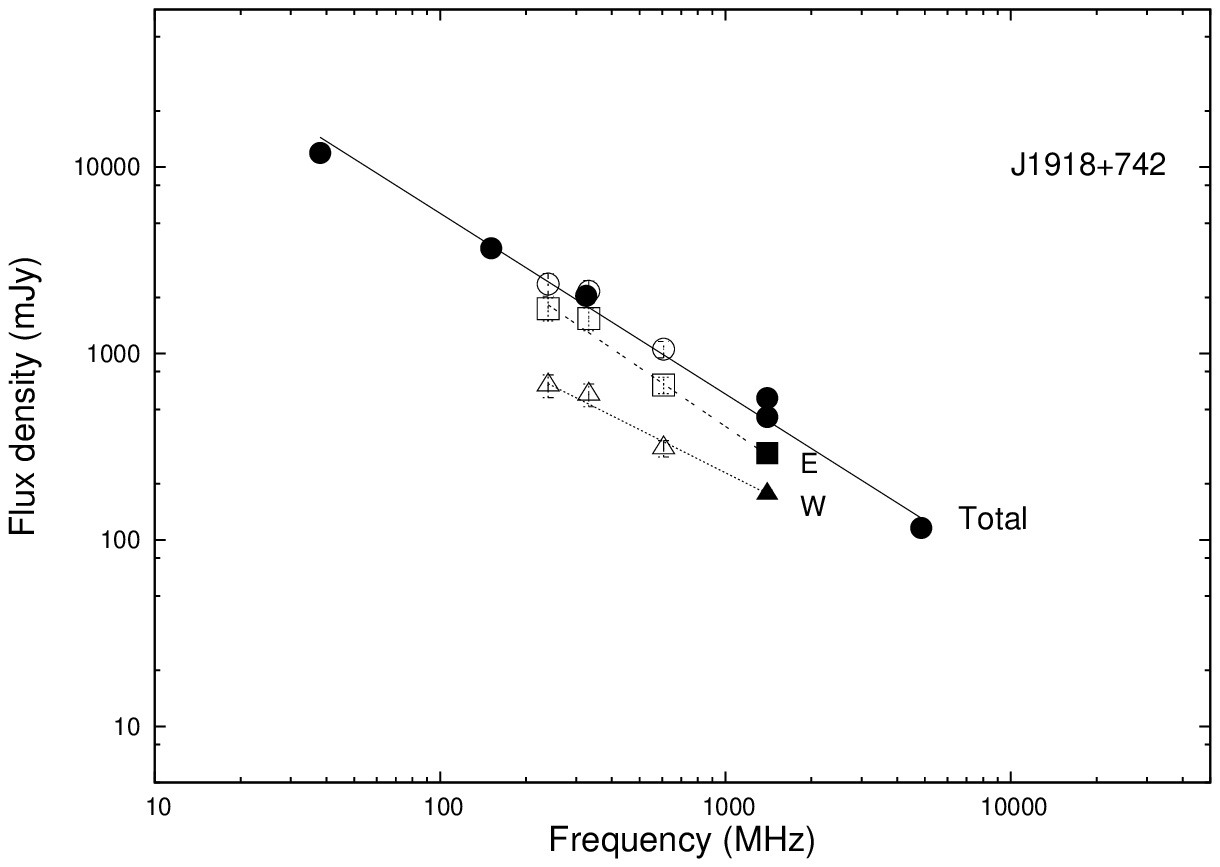} \qquad
            \includegraphics[width=6.2cm]{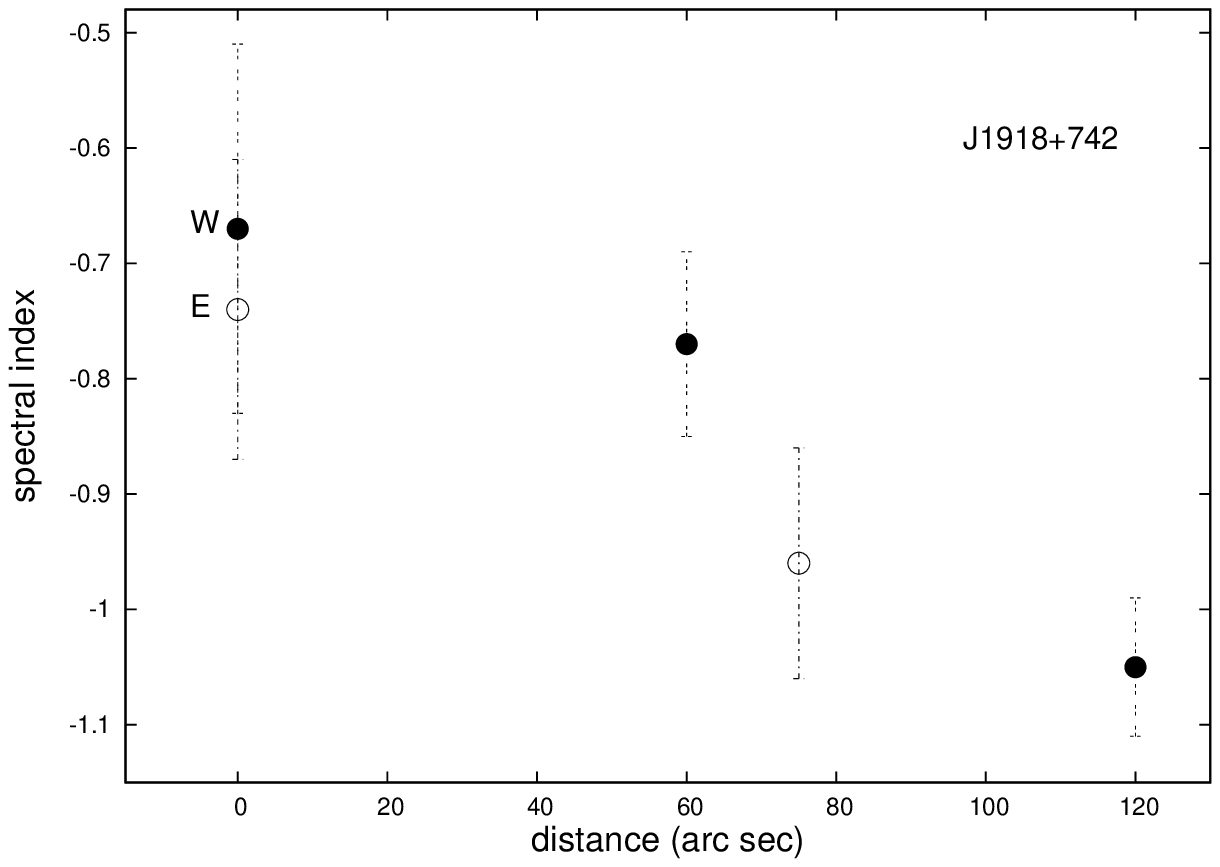}}
\caption{Left panel: the spectra of the J1918+742 and its components.
          The values have been taken from the following references: 38 MHz, Rees (1990); 151 MHz, Hales et al. (1991);
          325 MHz, Westerbork Northern Sky Survey (WENSS) survey, Rengelink et al. 1997;
          1400 MHz, NVSS survey; 1400 MHz, White \& Becker (1992); 4850 MHz,
          Gregory \& Condon (1991); 239, 332 and 607 MHz from this paper.
          Right panel: The spectral index profiles for the western and eastern components with an angular resolution
          of 45 arcsec using our measurements at 239, 332 and 607 MHz, and the NVSS image at 1400 MHz. For J1918+742
          one arcsec corresponds to 3.190 kpc.
        }
\end{figure}
%%%%%%%%%%%%%%%%%%%%%%%%%%%%%%%%%%%%%%%%%%%%%%%%%%%%%%%%%%%%%%%%%%%%%%%%%%%%%%%%%%%%%%%%%%%%%%%%%%%%%%%%%%

It is relevant to note that the frequency range used for estimating 
the integrated spectral indices is larger than that of the components, with total
flux density measurements being available at 38 MHz. Spectral indices at low frequencies,
especially at the outer edges of the lobes, 
reflect the injection spectra, and our values are consistent with earlier estimates 
in the literature.  For comparison, we consider a few other studies. 
Leahy, Muxlow \& Stephens (1989) estimated the spectral indices of hot-spots
between 151 and 1400 MHz, and found these to range from approximately $-$0.3 to $-$1.16 with a 
median value of approximately $-$0.8.  
Liu, Pooley \& Riley (1992) estimated the injection spectral indices for a sample of 
3CR FRII sources using flux density
measurements between 38 MHz and $\sim$1 GHz, and found that these values range from $-$0.65 to $-$1, with
a median value of $-$0.77. 
For a sample of giant radio
sources, the injection spectral indices were found to be in the range of $-$0.55 to $-$0.88, with a
median value of $-$0.6 (Konar et al. 2008; Jamrozy et al. 2008). However it is important to examine this 
further from observations at even lower frequencies by telescopes such as the LOw Frequency ARray (LOFAR).

High-frequency measurements are required to determine the break frequencies and spectral ages reliably. 
There is marginal evidence of spectral steepening in the lobes of J1211+743 from the 10500 MHz flux 
densities, but it would useful to confirm this from further observations at higher frequencies. The 
minimum energy magnetic field for the northern and southern lobes of J1211+743 are 0.15 and 0.14 nT
respectively, yielding spectral ages of $\sim$30 Myr for a break frequency of $\sim$10 GHz. The minimum 
energy magnetic field of the jet is 0.19 nT which yields an age of $\lapp$50 Myr for a break frequency 
$\gapp$5 GHz. The minimum energy magnetic field  estimates of western and eastern lobes of J1918+742 
are 0.16 and 0.20 nT respectively yielding ages of $\lapp$70 Myr for break frequencies $\gapp$1.4 GHz. 

%%%%%%%%%%%%%%%%%%%%%%%%%%%%%%%%%%%%%%%%%%%%%%%%%%%%%%%%%%%%%%%%%%%%%%%%%%%%%%%%%%%%%%%%%%%%%%%%%%%%%%%%%%

\section{Concluding remarks}
\label{sec:con_remk}

Although models of evolution of jets in an initially asymmetric environment in the
vicinity of the host galaxy lead us to expect that large radio sources should be
reasonably symmetric in both flux density and separation of the oppositely-directed
lobes from the parent galaxy (e.g. Jeyakumar et al. 2005), there are a number of highly 
asymmetric sources even
on Mpc scales. These could be due to either environmental asymmetries on Mpc scales 
caused by the filaments, sheets and voids, or intrinsic asymmetries in the jets. 
Some, such as the ones studied here, namely J1211+743 and J1918+742, have a diffuse
lobe without a significant hot-spot on one side of the parent galaxy, reminiscent of
the lobes of FRI galaxies, and an edge-brightened lobe with a hot-spot on the opposite
side. J1211+743 has a prominent jet with the magnetic field lines 
following the bends in the jet, as seen in the FRII sources, but points in the 
direction of the diffuse lobe.
This suggests that classification of hybrid morphology sources on the basis of a 
diffuse FRI-like lobe without a hot-spot may not be adequate.  
The asymmetries in brightness and structure 
of the oppositely-located components in these two large Mpc-scale sources are likely to 
be intrinsic rather than due to the effects of orientation and relativistic motion. The
diffuse lobe facing the jet in J1211+743 could be due to a highly dissipative jet, as has
been suggested earlier for several quasars without prominent hot-spots on the jet side.

The spectra of the lobes in J1211+743 and J1918+742 with a prominent peak or 
hot-spot tend to be flatter, and the spectral index appears to steepen with distance 
from the hot-spot, as has been seen earlier for most smaller-sized as well as  giant
radio sources. The variation
of spectral index in the diffuse northern lobe of J1211+743 is small
within $\sim$100 kpc of its outer edge, suggesting re-acceleration of particles.  
The integrated low-frequency spectral indices and also the values of $\alpha$ in the 
outer edges of the lobes are in the range of $\sim$$-$0.65 to $-$1.05, consistent with 
earlier low-frequency observations of luminous FRII radio sources. 
Observations at even lower frequencies by telescopes such as LOFAR,
would be valuable to further probe the injection spectral indices and constrain
particle acceleration models.

%%%%%%%%%%%%%%%%%%%%%%%%%%%%%%%%%%%%%%%%%%%%%%%%%%%%%%%%%%%%%
\section*{Acknowledgments}
We thank Gopal-Krishna, Chiranjib Konar and Paul Wiita for their comments and suggestions 
on an early version of this manuscript, and Dave Green for very detailed comments
on the present version of the manuscript. 
AP and SN thank NCRA, TIFR for hospitality during the course of this work, and
Kumaon University, Nainital, for registering them as PhD students.
AP is thankful to Jawaharlal Nehru Scholarship Ref: SU-A/287/2011-12/.
AP, SN and MS are also thankful to DST, Government of India for financial support
vide Grant No. SR/S2/HEP-17/2005.
The GMRT is a national facility operated by the National Centre
for Radio Astrophysics of the Tata Institute of Fundamental Research. We thank the staff
for help during the observations. The National Radio Astronomy Observatory  is a
facility of the National Science Foundation operated under co-operative
agreement by Associated Universities Inc. We are also thankful to the VLA staff for easy access
to the archival data base. This research has made use of the NASA/IPAC extragalactic database (NED)
which is operated by the Jet Propulsion Laboratory, Caltech, under contract
with the National Aeronautics and Space Administration. We thank numerous contributors
to the GNU/Linux group.
%%%%%%%%%%%%%%%%%%%%%%%%%%%%%%%%%%%%%%%%%%%%%%%%%%%%%%%%%%%%%%%%%
%------------------------------------------------------------------------------%
{}

%------------------------------------------------------------------------------%
\label{lastpage}
\end{document}